# Direct Measurement of the Electronic Structure and band gap nature of atomic-layer-thick 2H-MoTe$_2$

Wenjuan Zhao,[1,2,#] Xieyu Zhou,[3,#] Dayu Yan,[1,2,#] Yuan Huang,[1,*] Cong Li,[1,2] Qiang Gao,[1,2] HongtaoRong,[1,2] Yongqing Cai,[1,2] Eike F. Schwier,[4] Dianxing Ju,[5] Cheng Shen,[1,2] Yang Wang,[1,2] Yu Xu,[1,2] Wei Ji,[3] Youguo Shi,[1] Lin Zhao,[1] Lihong Bao,[1] Qingyan Wang,[1] Kenya Shimada,[4] Xutang Tao,[5] Hongjun Gao,[1,2] Zuyan Xu,[6] Xingjiang Zhou,[1,2,7,8,*] Guodong Liu[1,7*]

[1] Beijing National Laboratory for Condensed Matter Physics, Institute of Physics, Chinese Academy of Sciences, Beijing 100190, China
[2] University of Chinese Academy of Sciences, Beijing 100049, China
[3] Department of Physics, Renmin University of China, Beijing 100872, P.R. China
[4] Hiroshima Synchrotron Radiation Center, Hiroshima University, Higashi-Hiroshima, Hiroshima 739-0046, Japan
[5] State Key Laboratory of Crystal Materials, Shandong University, 250100, Jinan, Shandong, China
[6] Technical Institute of Physics and Chemistry, Chinese Academy of Sciences, Beijing 100190, China
[7] Songshan Lake Materials Laboratory, Dongguan, Guangdong 523808, China
[8] Beijing Academy of Quantum Information Sciences, Beijing 100193, China

## Abstract

The millimeter sized monolayer and bilayer 2H-MoTe$_2$ single crystal samples are prepared by a new mechanical exfoliation method. Based on such high-quality samples, we report the first direct electronic structure study on them, using standard high resolution angle-resolved photoemission spectroscopy (ARPES). A direct band gap of 0.924eV is found at $K$ in the rubidium-doped monolayer MoTe$_2$. Similar valence band alignment is also observed in bilayer MoTe$_2$，supporting an assumption of a analogous direct gap semiconductor on it. Our measurements indicate a rather large band splitting of 212meV at the valence band maximum (VBM) in monolayer MoTe$_2$, and the splitting is systematically enlarged with layer stacking, from monolayer to bilayer and to bulk. Meanwhile, our PBE band calculation on these materials show excellent agreement with ARPES results. Some fundamental electronic parameters are derived from the experimental and calculated electronic structures. Our findings lay a foundation for further application-related study on monolayer and bilayer MoTe$_2$.

## Introduction

Two-dimensional (2D) crystals, such as graphene [1], hBN [2,3], phosphorene [4], transition metal dichalcogenides (TMDCs) [5,6] and van der Waals (vdW) ferromagnets [7] etc., provide a unique platform for exploring novel physical properties and functionalities not existing in their bulk counterparts. Of particular interest is the TMDC $MX_2$ (M = Mo, W; X = S, Se, Te) semiconductor, which shows very distinct properties in the monolayer limit, including direct band gap transition [8-11], well-defined valley degrees of freedom and spin-valley locking [12], huge exciton binding energy [13-14], Ising superconductivity [15] and so on. Based on these unique properties of monolayer $MX_2$ materials, many application researches have been made for functional testing and device demonstration [16]. Besides the distinct physical properties, 2D TMDCs materials are compatible with modern semiconductor planar processing, which makes them show great potential in the next-generation nanoelectronics [16], optoelectronics [17] and valleytronics [14].

Among TMDCs, 2H-$MoTe_2$ is very intriguing with respect to basic physical properties and application. First, 2H-$MoTe_2$ has the largest lattice parameters among $MX_2$ semiconductor. Due to rather large interlayer distance, its three-dimensional character is rather weak. Second, in $MoTe_2$ depending on growth conditions, there exists three different structures: hexagonal 2H phase (semiconductor), monoclinic 1T' phase (metal) and orthorhombic $T_d$ phase (Weyl semimetal) [18,19]. An electrostatic-doping-driven phase transition, between hexagonal and monoclinic phase, has been reported to manipulate the electronic state [20]. Moreover, atomically thin $MoTe_2$ may be particularly promising for optoelectronics applications, considering its weak exciton–phonon interactions [21], very large spin–orbit coupling of 250 meV [22,23], and stable exciton state due to large bright-dark exciton splitting of 25 meV [24], as well as the large luminescence yield in bilayer $MoTe_2$ sheets [25-27]. However, up to now, there has been no electronic structure measurement being reported on monolayer and bilayer $MoTe_2$. .

In spite of being fundamentally important, detailed and systematic ARPES research on the electronic structure of mono-to-few-layer $MX_2$ crystals is quite scarce, which is mainly restricted by obtaining large-size, uniform-thickness and high-quality samples. For a few studies on exfoliated samples [28-30], the resolution and data statistics are quite limited, arising from tiny sample size and limitation of nano(micro)-ARPES technique. In another few studies on atomically thin $MX_2$ films grown by the sophisticated MBE or CVD method [31-35], the standard ARPES technique was employed to measure band structure. Such films contain large quantity of micrometer-sized grains though identically orientated.

In this letter, by using a new mechanical exfoliation method [36], we have succeeded in fabricating millimeter size of monolayer and bilayer $MoTe_2$ sample with high quality. With these samples, combining high resolution of ARPES, we present the first direct observation of low-energy electronic structure of monolayer and bilayer $MoTe_2$. Our work will contribute to develop scalable and high performance electronic devices based on atomically thin $MoTe_2$.

**Experimental**

Monolayer, bilayer and multilayer $MoTe_2$ flakes were prepared by a new mechanical exfoliation method from bulk crystal and transferred on to a $SiO_2$/Si wafer

[36]. A more detailed description of sample preparation is shown in the Supplemental Material [37]. The ARPES measurements on monolayer and multilayer samples were performed with photon energy of 21.218 eV, using our home-build Photoemission spectroscopy system [38] equipped with a VUV5000 Helium lamp and a DA30L electron energy analyzer (Scienta Omicron). The overall energy resolution was set at 10 - 20 meV, and the angular resolution was 0.2 degree. The bilayer samples were measured on BL-1 end station of Hiroshima Synchrotron Radiation Center (HiSOR) with photon energy of 45 eV. After being cleaned through mild vacuum annealing at ~500 K, all the samples were measured at low temperature of 30 - 40K and in ultrahigh vacuum with a base pressure better than $5 \times 10^{-11}$ mbar.

## Results and Discussion:

Fig. 1 shows the measured constant energy contours (CECs) in $\Gamma$-$K$ region and only around K region for a typical millimeter-sized monolayer MoTe$_2$ sample. From the upper row, it can be seen that the CEC is evolving from a spot shape to a two concentric hollow structure with the increased binding energy. In particular, until down to deep binding energy of 1.05eV, the CEC spectral weight appears just around $K$ but not around $\Gamma$. This is consistent with the direct gap feature of monolayer MX$_2$. The lower panels reveal the complete feature of CEC around $K$, showing an evolution from a round spot, to single circle and to concentric double triangles, when going into higher binding energy. Such triangular warping reflects the anisotropy of electronic states around $K/K'$, which stems from the three-fold rotational symmetry of the crystal structure [39].

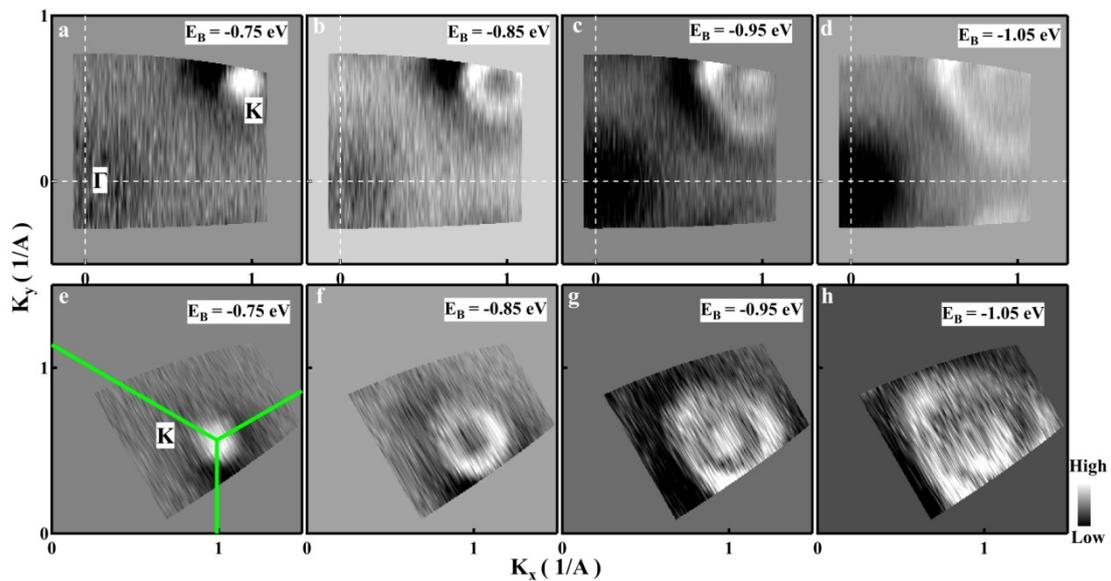

*Figure 1 Constant energy contours of exfoliated monolayer MoTe$_2$. (a)-(d) Constant energy slices crossing $\Gamma$-$K$ region. (e)-(h) Constant energy contours mapped around K. These mappings exhibit evolution of monolayer MoTe$_2$ valence band features*

*around Γ and K points. To eliminate the background effect on visibility, only second-derivative spectra are presented here. The data were collected at ~30 K with the photon energy of 21.2eV (He I line).*

Fig. 2 (a-f) gives an overview of the monolayer, bilayer and bulk MoTe$_2$ band structure along the *M-Γ-K-M* high symmetry lines of the Brillouin zone. The upper panels (a) to (c) show the original band dispersion obtained by matching the segmented data together. To enhance the visibility, the lower panels (d) to (f) present the second-derivative spectra of original data with respect to energy, which are overlaid with the calculated band structures as the red dotted lines. To match the measured band position, the calculated data are shifted in energy.

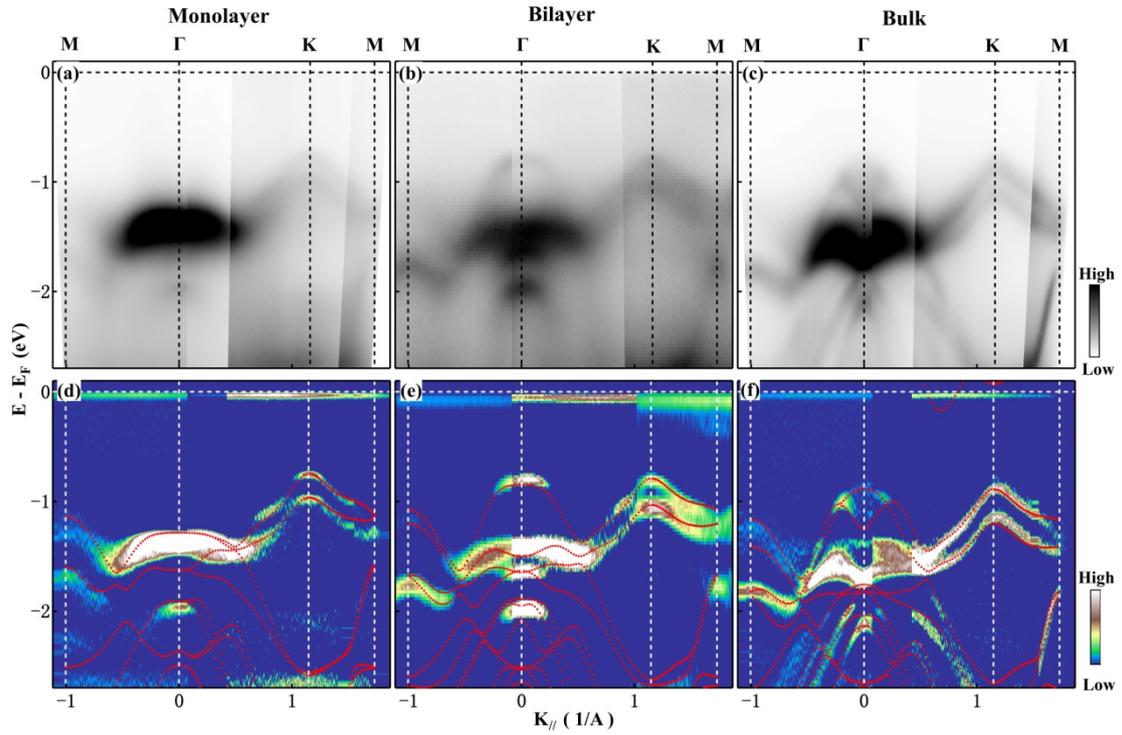

*Figure 2. Electronic band structure along M-Γ-K-M high symmetry line in monolayer (a), bilayer (b) and bulk (c) MoTe$_2$. The upper panels show the original ARPES spectra. The lower panels display the second derivative plots of these spectra with respect to energy. The red dotted lines are the corresponding calculated bands from our DFT-PBE calculation.*

By comparison, it is clear that the measured band structures for these samples are highly consistent with the band calculations with spin-orbit coupling taken into account. For the monolayer MoTe$_2$ sample, as shown in Fig. 2(a), the most remarkable band features include a long flat band centered at *Γ* and a rather identical band splitting around *K*. They are also common features among other monolayer MX$_2$ [31-35]. From *K* to *Γ*, two splitting bands merge into one degenerate band at about middle momentum position, namely, the flat band around *Γ*. The VBM at *K* is positioned at significantly lower binding energy (-0.76 eV) than that (-1.375 eV) at $\Gamma$ (Fig. 3(a) and 3(e)). In other words, the position of the topmost valence band (VB) at $K$ is much higher than that at $\Gamma$. Distinct from the monolayer sample, the repulsion between orbitals perpendicular to plane, i.e. $d_{z2}$ of Mo and $p_z$ of Te, increase the energy at $\Gamma$ in bilayer and bulk MoTe$_2$. However, this effect reduce from $\Gamma$ to $M$ and $\Gamma$ to $K$ because of the increasing component of in-plane orbitals. As a result, the top most valence band in the $\Gamma$ region is changed into a downwards parabolic-like band instead of a flat band. In these two cases, the VBM at $K$ are close to or slightly higher than that at $\Gamma$ [bilayer: -0.795eV ($K$) vs -0.795eV ($\Gamma$); bulk: -0.87eV ($K$) vs -0.665 ($\Gamma$) eV ], that is different from other typical MX$_2$ semiconductors MoS$_2$, WS$_2$ and MoSe$_2$, where VBM at $\Gamma$ is significantly higher than that at $K$ in both bilayer and bulk material [28,31,40-42]. However, here the situation in MoTe$_2$ is similar to the case in WSe$_2$ [32, 42,43]. References 32 and 43, reported that bilayer WSe$_2$ on a graphene substrate or after being strained could also be a direct gap semiconductor. Considering good agreements between calculated and measured bands, such kinds of observed band evolution provides firm support for the indirect- to-direct band gap transition in thinning MoTe$_2$ crystal from bulk down to few-layer and to monolayer finally, as evidenced in photoluminescence experiments [25-27,44]. This electronic state change originates from quantum confinement effect [9,45].

According to our band calculation [37], in the monolayer MoTe$_2$, the top most and long flat band around $\Gamma$ is derived mainly from Mo $d_{z2}$ orbital and little from Te $p_z$ orbital. Therefore this state manifests predominant localization in plane of lattice, resulting in full flatness of VBM. Around $K$, two splitting valence bands are contributed dominantly by Mo $d_{x2-y2}/d_{xy}$ orbitals and little from Te $p_x/p_y$ (bonding state), which are rather extended in plane, giving rise to clear dispersion of bands. While the splitting of bands is primarily induced by spin orbital coupling arising from Mo atoms due to the absence of structural inversion symmetry. For bilayer and bulk MoTe$_2$, the top-most valence bands at $\Gamma$ have the same orbital character of Mo $d_{z2}$ bonding state as in monolayer sample, but the second top-most valence bands VB2 are induced by Mo $d_{z2}$ and Te $p_z$ orbitals, respectively. Their two upper VBs around $K$ are all originated from Mo $d_{x2-y2}$ and $d_{xy}$ orbitals (bonding state) like in monolayer sample, but with degenerate spin states for each band, since the inversion symmetry is introduced in the bilayer and bulk. It will be discussed again in the following text.

Like other monolayer (Mo, W)(S, Se)$_2$ semiconductors with a direct band gap at $K$, the 2H-MoTe$_2$ monolayer is also a direct-gap semiconductor as evidenced by band calculation [26,37,46] and photoluminescence measurement [25-27]. To directly confirm such a band gap character, we need determine the conduction band minimum (CBM) location in the whole BZ. It is accessible for ARPES by filling electrons into the unoccupied conduction bands, through surface doping with alkaline metal atoms. With adequate Rb doping, we found that the CBM becomes visible only at $K$ below Fermi level (inset of Fig. 3(d)). Figs. 3(a)–3(h) present the band spectra of pristine and three cumulatively rubidium (Rb)-doped MoTe$_2$ monolayer along two different momentum directions of BZ, crossing adjacent $K$ and $K'$ respectively, as marked by blue lines (cut 1 and cut 2) in the inset of Fig. 3(a). For the pristine sample, as shown in Fig 3(a) and 3(e), the VBM along cut 1 and cut 2 is at $K'$ and $K$ with the same binding energy of 0.76 eV, respectively. With increasing Rb deposition, these bands shift progressively down towards higher binding energies. Along cut 1, the additional weak and near-$E_f$ spectral weight can be identified as the CBM at $K'$ (Fig. 3(d)), which is aligned with the VBM with respect to momentum. This reveals the direct electronic band gap in monolayer MoTe$_2$. Nevertheless, the spectral signature of CBM at $K$ is hard to be found in Fig. 3(h), due to the nearly disappeared matrix element. The gap value at $K'$ in Fig. 3(d) can be read out to be 0.924 eV, which is quite close to the gap size of 0.97 eV from our PBE calculation. However, this is much different from other monolayer MX$_2$ materials [31-33], in which the extracted gap energy from ARPES spectra was substantially reduced, comparing with the gap value determined by optical measurements and band calculation. We suppose it is mostly because the local work function of tellurium atoms in monolayer MoTe$_2$ may be rather close to the underneath gold, that makes nearly no charge transfer between sample and substrate and reduces the screening effect thereby.

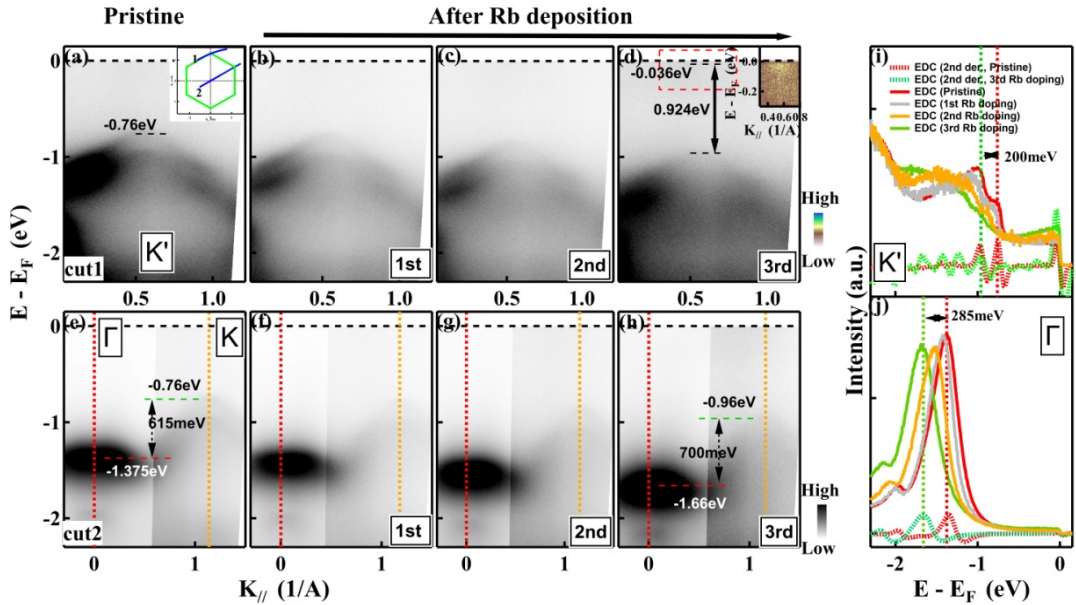

*Figure 3. Direct band gap observation through accumulative rubidium deposition on monolayer MoTe$_2$. The band dispersions of pristine (a) and Rb-doped (b)-(d) MoTe$_2$ monolayer, measured along cut 1 as shown in the inset of panel (a). Direct band gap is marked with black arrows in (d). (e) and (f)-(h) Same as in (a) and (b)-(d) but along cut 2 (i.e. Γ-K direction). The horizontal dash lines and numbers indicate the band shifting with Rb doping. (i) Energy distribution curves (EDCs) of the valence band at K' point. The red solid and dashed lines are extracted from the original (a) and its second-derivative spectra of pristine sample, respectively. Similarly, the grey, orange and green solid and green dashed lines are for the Rb-doped sample (b)-(d). (j) EDCs of the valence band in (e-h) at Γ point. The curves' definition is the same as in panel (i). The momentum positions for taking EDCs are marked by vertical dash lines in (e-h).*

What's more, to quantitatively evaluate the band shifting with Rb doping, we plot the EDCs taken from the original band spectra and their second-derivative plots in Fig. 3(i) at $K'$ and Fig. 3(g) at $\Gamma$. After the third Rb deposition, it can be seen that the VBM at $K'$ shifts down by 0.96-0.76=0.2 eV upon electron doping, while band maximum at $\Gamma$ moves down more by 1.66-1.375=0.285eV. It means that the band shifting at $K'$ is smaller than that at $\Gamma$, which reveals a non-rigid band shift with rubidium doping in monolayer MoTe$_2$. It provides a possibility of engineering electronic structure by surface doping of electrons in monolayer MoTe$_2$.

The up-and-down band splitting around $K$ is related to many important phenomena and properties in MX$_2$ materials, such as direct gap transition, spin-valley locking, AB exciton effect [13,14] and spin-layer locking [47]. In Fig. 4, we show this band splitting along $\Gamma$-$K$ direction with the exactly same momentum range, for monolayer, bilayer and bulk samples. We can see the sharp band dispersion of the original ARPES spectra and their second-derivative plots shown in Fig. 4(a-c) and (d-f), respectively. As shown in Fig. 4(g-i), the quantitative band splitting at $K$ can be extracted from the respective EDCs for the original bands (red curves). By Gaussian fitting curves shown as blue dotted line [37], we can clearly see that the splitting size at $K$ is systematically enlarged with layer stacking, from 212 meV in monolayer, to 225 meV in bilayer and to 252 meV in bulk MoTe$_2$. Such observed splitting data show very good agreement with our PBE calculation of 220 meV, 240meV and 310meV for these samples, respectively [also see table I].

Besides gap, band splitting at $K$, and VBM difference from $K$ to $\Gamma$, other electronic parameters like effective mass can be easily extracted by fitting to the measured and calculated band structure. We summarized and listed all these electronic parameters in table I, where $K$ ($\Gamma$) effective mass (VB1) denotes the electron effective mass derived from the top most valence band at $K$ ($\Gamma$) along $\Gamma$-$K$ direction. From monolayer to bilayer and bulk, the observed hole effective mass at $K$ for VB1 increases firstly from 1.17 to 1.69 m$_e$ and then decreases down to 1.56 m$_e$. This variation is on the trend consistent with our PBE calculation. While at the center of Brillouin zone, both observed and calculated hole effective mass for VB1 shows fast and monotonic decreasing with accumulated layer numbers.

For bilayer MoTe$_2$, our measurement indicates that the VBM at $K$ is of same binding energy (-0.795 eV) as that at $\Gamma$ (also -0.795 eV). While our PBE calculation predicts the former is only a little higher than the latter in energy, signaling the VBM still being at $K$ point. This is different from other Mo-based bilayer semiconductors Mo(S, Se)$_2$, where the VBM shifts from $K$ to $\Gamma$. Our finding might be consistent with conclusion of the earlier reported photoluminescence (PL) experiments [25-27], suggesting the same direct gap nature of bilayer MoTe$_2$ as in its monolayer version..

In general, in the MX$_2$ family of materials, the band splitting of VBM around the $K$ point is caused by spin-orbit coupling, inversion symmetry breaking and interlayer coupling [12]. Spin-orbit coupling (SOC) is a very fundamental and profound interaction in MX$_2$, which stems from a rather large intra-atomic spin-orbit interaction of the transition metal M atoms. The SOC induced band splitting is usually determined by SOC of the atoms themselves and intra-layer inversion symmetry breaking in 2H phase TMDs. For monolayer $MX_2$, a spin splitting at $K$ is concomitantly observed [12,48,49]. This splitting disappears in the $\Gamma-M$ lines because of time reversibility [50]. The spin splitting size (212 meV) of $MoTe_2$ at $K$ is largest among three monolayer Mo-based $MX_2$ semiconductors, with other two of $MoSe_2$ (~180 meV [31]) and $MoS_2$ (~150 meV [33]). Considering the enhanced magnitude of spin–orbit coupling with the increase of atomic number, it is a natural consequence. This makes monolayer $MoTe_2$ more suitable to be used in spintronics application.

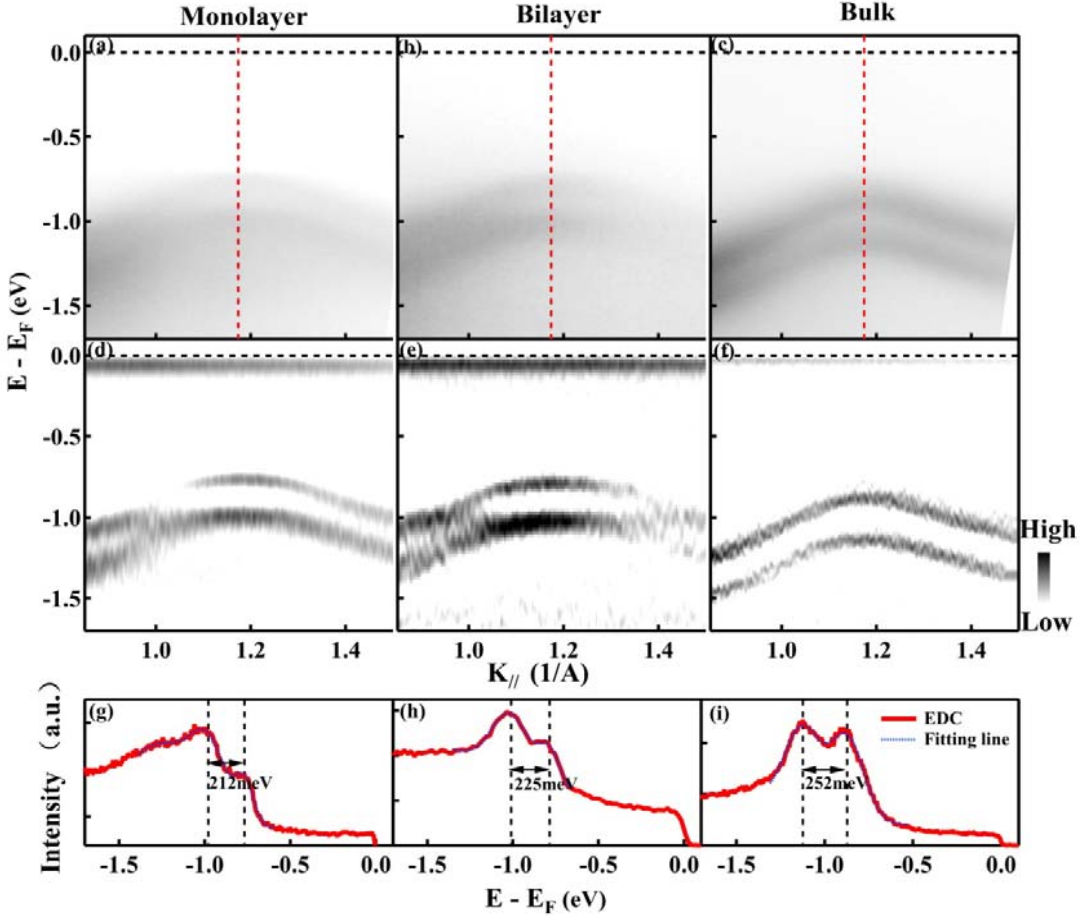

*Figure 4. Layer dependent valence band splitting at K. (a-c) The measured band splitting around K along Γ-K-M direction for monolayer, bilayer and bulk MoTe$_2$, respectively. (d-f) Second-derivative spectra corresponding to the data in (a-c). (g-i) The EDCs of the split valence bands at K points, shown as red solid lines. The blue dotted lines are corresponding to their multi-peak Gaussian fitting. The extracted splitting sizes are labeled by the double-headed arrows and numbers.*

*Table 1: Electronic parameters for the PBE calculated and ARPES measured monolayer, bilayer and Bulk MoTe$_2$. The units of parameters are marked in []. The band positions of VB1 are shown in Fig. S4.*

| Layers | Monolayer | | Bilayer | | Bulk | |
|---|---|---|---|---|---|---|
| | ARPES | Calculation | ARPES | Calculation | ARPES | Calculation |
| **K splitting [eV]** | 0.212 | 0.22 | 0.225 | 0.24 | 0.252 | 0.31 |
| **K-Γ VBM [eV]** | 0.615 | 0.536 | 0 | 0.059 | -- | -0.005 |
| **Gap size [eV]** | 0.924 | 0.97 | -- | 0.91 | -- | 0.72 |
| **K effective mass (VB1) [$m_0$]** | 1.17 | 0.59 | 1.69 | 0.62 | 1.56 | 0.61 |
| **effective mass (VB1) [$m_0$]** | 13.09 | 34.52 | 5 | 8.5 | -- | 1.27 |

In bilayer and bulk $MoTe_2$, there exist both the space inversion symmetry [$E_\downarrow(k) = E_\downarrow(-k)$] and time inversion symmetry [$E_\downarrow(k) = E_\uparrow(-k)$]. Hence, spin degeneracy $E_\downarrow(k) = E_\uparrow(k)$ must present in reciprocal space when having no external magnetic field. The calculated band structure of bilayer and bulk $MoTe_2$ without including SOC are shown in Fig. S5. The splitting at K still exists though no SOC is considered, suggesting its origin of interlayer coupling. When SOC is introduced, band VB1 and VB2 split into two bands, VB1a and VB1b, VB2a and VB2b, respectively. However, the interlayer inversion symmetry leads to the degeneration of VB1a-VB2a and VB1b-VB2b. It is considered that the merely SOC-induced band splitting size around $K$ is comparable in size among monolayer, bilayer and bulk $MoS_2$ [12]. Consequently, given that this fact is ubiquitous in all $MX_2$ family, we can attribute the difference of band splitting size to different interlayer coupling strength between bilayer (225 meV), bulk (252 meV) and monolayer (212 meV) $MoTe_2$. Thus, our observation gives an interlayer interaction strength 13 meV and 40 meV for bilayer and bulk $MoTe_2$, respectively. From bilayer to bulk, the increased interlayer coupling strength should come from layer separation reducing after increasing number of layers that enhance the interlayer hopping and coupling in turn [51]. In spite of many studies, there has been no specific ARPES study to quantify the valence band splitting in few layer $MX_2$, although it ought to be the most direct tool. Our study paves a successful way to directly and quantitatively investigate the interlayer coupling for the few-layer $MX_2$. Also it is significant to understand and clarify the origin of valence band splitting, since the unique spin and valley physics in 2D $MX_2$ crystal is just governed by such a splitting.

## Conclusion:

In summary, by taking high-resolution ARPES measurements and performing ab-initio band structure calculations, we have systematically studied the electronic structures of large-area exfoliated monolayer and bilayer $MoTe_2$. Our study provides the first direct experimental evidence of direct gap semiconductor for monolayer $MoTe_2$. We also observe that the bilayer $MoTe_2$ has the similar valence band alignment as its monolayer counterpart. It suggests a possible direct band gap nature for bilayer $MoTe_2$ either. The measured valence band splitting at $K$ is found to systematically increase with layer stacking. With these data, we can separately identify the interlayer coupling and spin-orbital coupling strength for bilayer and bulk

MoTe$_2$. Surface doping on monolayer MoTe$_2$ by alkaline metal Rb induces a non-rigid band shift, pointing to importance of electron correlation in the process of carrier doping. In addition, our experimental results show very good agreement with the band structure calculations on both monolayer and bilayer as well as bulk MoTe$_2$. Our study pave a way to thoroughly understand electronic properties and develop scalable and high performance electronic devices based on atomic layer thick MoTe$_2$.

## Acknowledgements

This work is supported by the National Science Foundation of China (Nos. 11574367 and 11874405), the National Key Research and Development Program of China (No. 2016YFA0300600), and the Youth Innovation Promotion Association of CAS (Nos. 2017013 and 2019007).

## Corresponding Authors:

yhuang01@iphy.ac.cn,　xjzhou@iphy.ac.cn,　gdliu_arpes@iphy.ac.cn